\documentclass[12pt]{article}
\usepackage{amsmath} 
\usepackage{graphicx} 
\usepackage{epsfig}
\usepackage{subfig} 
\usepackage{float} 
\usepackage{longtable}
\usepackage{times}
\usepackage{txfonts}
\usepackage{color}
\usepackage{ulem}
\usepackage[babel=true]{csquotes}          

\usepackage{lineno}

 \usepackage{ifpdf}
 \ifpdf
 \usepackage[pdftex]{hyperref}
 \else
 \usepackage[hypertex]{hyperref}
 \fi

 \hypersetup{
   pdftitle={},%
   pdfauthor={},%
   pdfsubject={},%
   pdfkeywords={},%
   pdfstartview={},%
   bookmarksopen=true, breaklinks=true, debug=true, %
   colorlinks=true, linkcolor=red, citecolor=blue, urlcolor=blue
 }

\newcommand{\beq}{\begin{eqnarray}}
\newcommand{\eeq}{\end{eqnarray}}
\newcommand{\be}{\begin{eqnarray*}}
\newcommand{\ee}{\end{eqnarray*}}

\newcommand{\eqs}[1]{\begin{equation} \begin{split} #1\end{split} \end{equation} }

\newcommand{\ie}{{\it i.e.}}
\newcommand{\eg}{{\it e.g.}}

\newcommand{\cf}[1]{{Fig.~\ref{#1}}}
\newcommand{\ct}[1]{{Table.~\ref{#1}}}

\def\lsim{\raise0.3ex\hbox{$<$\kern-0.75em\raise-1.1ex\hbox{$\sim$}}}
\def\gsim{\raise0.3ex\hbox{$>$\kern-0.75em\raise-1.1ex\hbox{$\sim$}}}

\def\dAu  {$d$Au}
\def\dAum  {d\mathrm{Au}}

\def\Ncoll   {\mbox{$N_{\rm coll}$}}

\def\jpsi    {\mbox{$J/\psi$}}
\def\upsi    {\mbox{$\Upsilon$}}

\def\beq     {\begin{equation}}
\def\eeq     {\end{equation}}

\long\def\symbolfootnote[#1]#2{\begingroup%
  \def\thefootnote{\fnsymbol{footnote}}\footnote[#1]{#2}\endgroup}

\begin{document}


\begin{center}
{\Large \bf {\boldmath \upsi} production in {\boldmath$p(d)A$} collisions at RHIC and the LHC}

\vskip 8. truemm
{\bf E. G. Ferreiro$^a$, F. Fleuret$^b$, J.P. Lansberg$^c$, N. Matagne$^d$, A. Rakotozafindrabe$^e$}\\
\vskip 5. truemm

{\it $^a$Departamento de F{\'\i}sica de Part{\'\i}culas and IGFAE, Universidad de Santiago de Compostela, 15782 Santiago de Compostela, Spain}\\ ~ \\
{\it $^b$Laboratoire Leprince Ringuet, \'Ecole Polytechnique, CNRS/IN2P3,  91128 Palaiseau, France}\\~ \\
{\it $^c$IPNO, Universit\'e Paris-Sud, CNRS/IN2P3, F-91406, Orsay, France} \\~ \\
{\it $^d$Universit\'e de Mons, Service de Physique Nucl\'eaire et Subnucl\'eaire, Place du Parc 20, B-7000 Mons, Belgium}\\~ \\
{\it $^e$IRFU/SPhN, CEA Saclay, 91191 Gif-sur-Yvette Cedex, France}

\vskip 5 truemm
\end{center}

\begin{abstract}
We study the effect of nuclear matter in $\Upsilon$ production in $d$Au collisions at RHIC
and $p$Pb collisions at the LHC. We find that the nuclear modification factor, $R^{\Upsilon}_{\dAum}$,  
measured at RHIC is not satisfactorily reproduced by the conventional effects used in the literature, namely
the modification of the gluon distribution in bound nucleons and an --effective-- survival probability for a bound state
to escape the nucleus. 
In particular, we argue that this probability should be close to 1 as opposed to the
$J/\psi$ case. 
We note that, at backward rapidities, 
the unexpected suppression of $R^{\Upsilon}_{d\rm Au}$ observed by PHENIX hints at the presence of a 
gluon EMC effect, analogous to the quark EMC effect -- but likely stronger.
Further nuclear matter effects, such as saturation and fractional energy loss, are
discussed, but none of them fit in a more global picture of quarkonium production.
Predictions for $ \Upsilon(nS)$ for the forthcoming $p$Pb run at 5~TeV
at the LHC are also presented.
\end{abstract}




\section{Introduction}
Quarkonium bound states, especially the \upsi 's, offer a solid ground to
probe the short distances behaviour of Quantum Chromodynamics (QCD), due
to the relatively high scale provided by the large mass of the heavy quarks.
In addition to the production mechanisms in the
vacuum~\cite{Lansberg:2006dh,Brambilla:2010cs}, the properties of production
and absorption of quarkonium in a nuclear medium~\cite{Brambilla:2010cs,Rapp:2008tf}
provide quantitative inputs for the study of Cold Nuclear Matter (CNM) effects
in proton-nucleus collisions and for the understanding of QCD at high density
and temperature in nucleus-nucleus collisions.

We show that the conventional nuclear modifications of the gluon distribution in 
heavy ions -- known as shadowing and anti-shadowing -- as well as the 
possible break up of the $b \bar b$ pair along its way off the nucleus 
are shown to have  a limited impact on the \upsi\ production in \dAu\ collisions at RHIC at 
$\sqrt{s_{{NN}}}=200 \,\mathrm{GeV}$.
Without additional effects, the nuclear modification of the
yield  available from the 
PHENIX experiment~\cite{Adare:2012bv} in the backward region is not satisfactorily reproduced.

This motivated us to study in detail other effects: (a) the impact of
Fermi motion on the gluon distribution in nuclei from unity down to rather low $x$ $\simeq 0.2$, (b)
 a suppression of the gluon distribution in nuclei for intermediate $x$, $0.35 \leq x \leq 0.7$, 
analogous to the quark EMC effect~\cite{Aubert:1983rq}, unobserved until now, (c) the possible
effect of the saturation of gluon dynamics at low $x$ and finally (d)  a  fractional energy 
loss~\cite{Arleo:2010rb} proportional to the projectile {parton} energy {and} caused by medium-induced radiations 
associated to the quarkonium hadroproduction provided that the heavy quark pair remains in a 
coloured state for some time\footnote{As we argue later, such effect would be in contradiction with the satisfactory~\cite{Lansberg:2012ta} description
of low $P_T$ $pp$ data by the Colour-Singlet Model~\cite{Lansberg:2006dh,Brambilla:2010cs}.}.

We have compared our results --from our Monte-Carlo framework {\sf JIN}~\cite{Ferreiro:2008qj}-- 
to the \dAu\ \upsi\ data available from RHIC experiments~\cite{Adare:2012bv,Reed:2010zzb} and 
we have predicted the trend of the nuclear modification factor for the forthcoming $p$Pb run at the LHC only using the
effect of nPDF. We have then discussed whether gluon saturation and fractional energy loss could
be applied within a coherent picture of $\Upsilon$ production.

The structure of the paper is as follows. In section 2, we discuss the propagation of
the (pre-resonant) $\Upsilon$ state in the nuclear matter and we explain why 
we believe that its survival probability is close to unity. In section 3, 
we discuss all the possible nuclear modifications of the gluon-momentum distribution in 
a nucleon embedded in a large nucleus which can impact on $\Upsilon$ production in $p$A 
collisions at RHIC and LHC energies. 
In section 4, we present our results for $d$Au collisions at RHIC
and compare with existing data. In section 5, we briefly sketch the expected trend for the  $\Upsilon$
nuclear modification factor in $p$Pb collisions at 5 TeV at the LHC for a mild
gluon shadowing and antishadowing. In section 6, we discuss on the possibility  of the presence 
of a fractional energy loss at forward rapidities. Section~7 gathers our conclusions.


\section{Propagation in cold nuclear matter: \upsi\ vs \jpsi}
 The first effect to be discussed is the probability for the heavy-quark pair to survive 
the propagation through the nuclear medium, 
usually parametrised by an effective cross section~$\sigma_{\mathrm{eff}}$. It is sometimes referred 
to as the nuclear absorption or break-up probability.

A priori, the smaller size of the $b{\bar b}$ pair when compared to $c{\bar c}$ pair implies that $b{\bar b}$ states 
should suffer less break-up than $c{\bar c}$. Yet, the ratio of their size depends on the evolution stage of the 
heavy-quark pair. At the production time, this ratio is expected to be $m_b/m_c$, hence a size 3 times smaller for 
$b \bar{b}$ than for $c \bar{c}$. When they are fully formed, it is rather 
$\frac{\alpha_s(2m_b)}{\alpha_s(2m_c)} \times  \frac{m_b}{m_c}$ as expected from their Bohr radii, hence a size 
2 times smaller. The relevant timescale to analyse the pair evolution is its formation time. 
According to the uncertainty principle, it is related 
to the time needed -- in their rest frame -- to distinguish the energy levels of the $1S$ and $2S$ states 
\cite{ConesadelValle:2011fw}:
$t_f=  \frac{2 M_{b\bar b}}{(M^2_{2S}-M^2_{1S})}=  2 \times 10$ GeV / 10.5 GeV$^2= 0.4$ fm for the \upsi.

For our purpose, $t_f$ has to be considered in the rest frame of the target nucleus, \ie~the Au beam at RHIC. 
The relevant $\gamma$ factor is then obtained from the rapidity of the pair corrected by the Au beam rapidity 
$\gamma=\cosh(y-y_{beam}^{\rm Au})$ where $y_{beam}^{\rm Au}=-5.36$ for RHIC. At the LHC, for a lead beam 
of 1.57 TeV in $p$Pb mode, $y_{beam}^{\rm Pb}=-8.11$. The formation time for the different 
rapidities reached by RHIC and LHC experiments are given in~\ct{tab:tf-RHIC} and~\ct{tab:tf-LHC}. 
$t_f$ is significantly larger than the Au and
Pb radii -- except in the most backward region. This implies that the $b {\bar b}$ is nearly always in a 
pre-resonant state when traversing the nuclear matter in both experimental set-ups. 
\begin{table}[htb!]
\begin{center}\setlength{\arrayrulewidth}{1pt}
\begin{tabular}{cccc|cccc}
\hline\hline
 $y$ & $\gamma(y)$ & $t_f(y)$   & \quad& & $y$  & $\gamma(y)$ & $t_f(y)$\\
\hline
-2.0 & 14.4     & 5.8 fm &   &   &  0.0 & 106      & 42 fm\\
-1.5 & 23.7     & 9.5 fm &   &   & +1.5 & 476      & 190 fm\\
-1.0 & 39       & 16 fm &   &   & +2.0 & 786      & 310 fm\\
\hline\hline
\end{tabular}
\caption{Boost and formation time in the Au rest frame of the $\Upsilon$ as a function of its rapidity
at $\sqrt{s_{NN}}=200$ GeV.}\label{tab:tf-RHIC}
\end{center}\vspace*{-0.5cm}
\end{table}

\begin{table}[htb!]
\begin{center}\setlength{\arrayrulewidth}{1pt}
\begin{tabular}{ccc|ccc|ccc}
\hline\hline
 $y$ & $\gamma(y)$ & $t_f(y)$    &  $y$  & $\gamma(y)$ & $t_f(y)$  &  $y$  & $\gamma(y)$ & $t_f(y)$\\
\hline
-4.0 & 20     & 8    fm &      -0.5 & $10^3$            & $4.0\times 10^2$   fm    &   1.5 & $7.5\times 10^3$   & $3.0\times 10^3$  fm \\
-3.5 & 50     & 20   fm &       0.0 & $1.7\times 10^3$  & $7.0\times 10^2$   fm    &   2.5 & $2.0\times 10^4$   & $8.0\times 10^3$ fm \\
-2.5 & 140    & 60   fm &       0.5 & $2.7\times 10^3$  & $1.1\times 10^3$ fm    &   3.5 & $5.5\times 10^4$   & $2.2\times 10^4$ fm \\
-1.5 & 370    & 150  fm &           &                   &                        &   4.5 & $1.5\times 10^5$   & $6.0\times 10^4$ fm \\
\hline\hline
\end{tabular}
\caption{Boost and formation time in the Pb rest frame of the $\Upsilon$
at $\sqrt{s_{NN}}=5$ TeV in $p$Pb collisions ($E_N^{\rm Pb}=1.57$ TeV and the Pb has a negative rapidity).}\label{tab:tf-LHC}
\end{center}\vspace*{-0.5cm}
\end{table}

For the forward and mid rapidity regions, this has two consequences: 
first, the break-up probability is expected to be small, of the order 
of a tenth of that of $J/\psi$ ($(m_c/m_b)^2 \sim 0.1$), following the 
early-time scaling $m_b/m_c$; and second, it ought to be the same for 
the \upsi($1S$), \upsi($2S$) and \upsi($3S$) states, since these 
states cannot be distinguished at the time they traverse the nucleus.
At the LHC (\ct{tab:tf-LHC}), $t_f$'s are even larger in the mid and forward-rapidity regions; the relative
suppression of the excited 
$\Upsilon$ states in PbPb collisions seen by CMS~\cite{Chatrchyan:2011pe} 
can thus only be explained by hot nuclear effects.

The backward region at RHIC as well as the most backward one at the LHC require a closer look. 
Indeed, for the same $t_f$, of the order of 15~fm, the E866 experiment at 
Fermilab \cite{Leitch:1999ea} observed a different suppression of 
\jpsi\ and $\psi(2S)$ produced with Feynman $x_F$ up to 0.2 at 
$\sqrt{s_{{NN}}}=38.8\,\mathrm{GeV}$. We might thus expect 
different absorption cross-section for the \upsi($1S$), \upsi($2S$) 
and \upsi($3S$) states for $y<-1$. However, the E772 experiment 
at Fermilab \cite{Alde:1991sw} has measured the 
\upsi($1S$) and \upsi($2S+3S$) separately at $\sqrt{s}=38.8$~GeV 
down to negative $x_F$ -- with even smaller $t_f$  -- and 
it observed a similar suppression for the $1S$ and the ${(}2S+3S{)}$ 
states\footnote{As it was discussed later on (see \eg~\cite{Leitch:1999ea}),
the E772 experiment suffered from a $P_T$ dependent acceptance,
especially in the backward region. For instance, the $J/\psi$
suppression was subsequently shown to be less marked than initially
thought. Yet, such a correction should equally apply for the 3
$\Upsilon$ states \cite{LeitchPriv}.}.
The only explanation for 
such a result is that the absorption of the bottomonium is 
actually very small, preventing us to see any difference of absorption 
between the 3 states. In the following, {we will consider a 
range of $\sigma_{\mathrm{eff}}$ from 0 to 1~mb} even though 1 mb has 
to be seen as a conservative upper bound.


\section{Gluon momentum distribution in the nucleus}
\subsection{Gluon shadowing and antishadowing}
At high energy (small $x_B$), 
the nuclear Parton Distribution Functions (nPDF) differ from those of free nucleons due to non-linear QCD effects. 
Nucleons {\it shadow} \cite{Glauber:1955qq,Gribov:1968jf} each other and the nPDFs are expected to be lower than for 
free nucleons.  
At $0.01 \leq x_B \leq 0.3$, some experimental data hint~\cite{Gousset:1996xt} at an excess of partons with regards to 
unbound nucleons, referred to as anti-shadowing. For $0.35 \leq x_B \leq 0.7$, the distribution is depleted again. This suppression 
is known as the EMC effect. 

These nuclear modifications are usually expressed in terms of the ratios $R_i^A$ 
of the nPDF of a nucleon bound in a nucleus~$A$ to the free nucleon PDF. The 
numerical parametrisation of $R_i^A(x_B,Q^2)$ is usually given for all parton 
flavours. Here, we limit our study to gluons since, at RHIC and the LHC, \upsi\ is essentially 
produced through gluon fusion~\cite{Lansberg:2006dh,Brambilla:2010cs}. To best explore 
the possible impact of $R_i^A$, we have considered 3 sets: EKS98~\cite{Eskola:1998df}, 
EPS08~\cite{Eskola:2008ca} and nDSg~\cite{deFlorian:2003qf} at LO. 
Recently, a new 
set with fit uncertainties, EPS09~\cite{Eskola:2009uj}, has been made 
available. Yet, in the case of gluons,  nDS and EPS08 match --except for $x_B \geq 0.3$-- the extreme values of
EPS09LO\footnote{Note, however, that the central curves for the LO and NLO EPS09 fits are different. This difference
is particularly large at low $Q^2$.}. Besides EKS98 is very close to EPS09LO central values. 
We thus consider more illustrative to use EKS98, EPS08 
and nDSg. The spatial dependence of the PDF nuclear modification has been included with a modification 
proportional to the local density~\cite{Klein:2003dj}.
Following the common practice, we label $x_1$ ($x_2$)
the gluon momentum fraction in the proton/deuteron (nucleus).

To account for the nuclear effects on \upsi\ production in nucleus collisions, 
we use our  Monte-Carlo framework {\sf JIN}~\cite{Ferreiro:2008qj}, based on the probabilistic 
Glauber model, used to describe \jpsi\ production at RHIC~\cite{OurExtrinsicPaper,Ferreiro:2009ur}.
It allows to consider improved kinematics corresponding to a $2\to2$ ($g+g\rightarrow b \bar{b}+g$) 
partonic process for the \upsi\ production (as in the Colour-Singlet Model (CSM) at LO~\cite{Brodsky:2009cf},
but also at higher-orders in $\alpha_S$ with 2 or 3 coloured partons in the final
state~\cite{Campbell:2007ws,Artoisenet:2008fc}).
In earlier studies of nuclear matter effects on \upsi\ production~\cite{Vogt:2010aa}, the $b \bar b$ pair was assumed to 
be produced by a $2\to 1$ partonic process ($g+g\rightarrow b \bar{b}$).
It would only apply
if Color-Octet Mechanism (COM) at LO were the relevant production
 mechanism at low transverse momenta. This is
disfavoured in view of the recent comparisons between the CSM predictions and the experimental 
data \cite{Lansberg:2012ta}, which leave little room for any COM contributions for $P_T \lesssim 5$~GeV.


\subsection{The EMC effect and the gluons} 
Thirty years ago, the European Muon Collaboration (EMC)~\cite{Aubert:1983rq}
observed a depletion of the quark densities in nucleons bound in nuclei, when 
compared to the ones of  free nucleons, in the range  of $0.35 < x_B < 0.7$.
To date, there is no consensus \cite{Norton:2003cb} about the origin of this suppression,
referred to as the EMC effect. It is still the subject of vivid activities. It has been 
attributed to local nuclear density effects, to properties of the bulk nuclear system, 
and recently \cite{Weinstein:2010rt} to Short Range Correlations (SRC) in the nucleus.
Up to now, this effect has not been {confirmed} for gluons,
even if it is allowed in some of the shadowing fits.

\begin{figure*}[htb!]
\begin{center}
\includegraphics[width=0.7\textwidth]{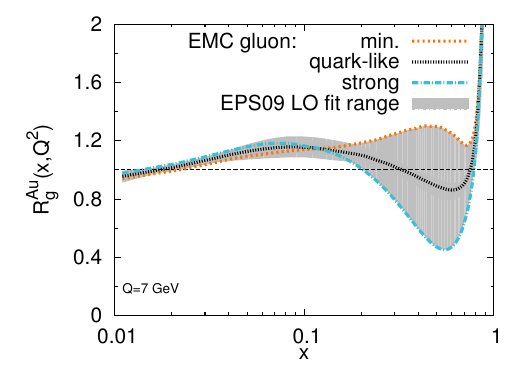}
\end{center}
\caption{The EPS09 uncertainty for the gluon density for Au at mid and large $x$ and 
the 3 gluon nPDF sets we have used to single out the
EMC effect.\label{fig:EPS09-EMC-range}}
\end{figure*}

While gluon shadowing in the existing constrained fits of nPDFs, especially
in EPS~08 \&~09, is the subject of intense on-going debates, the gluon EMC suppression 
is usually overlooked. Indeed, very little is known about gluons in this region and few data 
constrain their distribution at $x_B$ larger than 0.3. The amount of the gluon EMC suppression is 
actually pretty much unknown~\cite{Eskola:2009uj} (see \cf{fig:EPS09-EMC-range}), except for a loose constrain set by 
momentum conservation. We also note that another effect \cite{Kopeliovich:2005ym} arising from 
momentum conservation within the nucleons could be at play at larger $x$, $x\geq 0.7$, \ie\ at larger $y$ 
or smaller $\sqrt{s_{{NN}}}$, and {hence} it is not applicable here.

In the following discussion, in order to single out a possible impact of the gluon EMC 
suppression on the \upsi~production at RHIC, 
we will use three of the EPS09 LO sets: one with a quark-like EMC 
gluon suppression, and the two limiting curves in the region 
$0.35 < x_B < 0.7$ (\cf{fig:EPS09-EMC-range}). 
As the data-theory comparison will show, its magnitude indeed seems stronger than what
has been previously supposed in most of the existing nPDF sets (which
assumed a quark-like EMC gluon suppression) 
and disagrees with a Fermi-motion enhancement down to $x_B \simeq 0.5$ as expected
in~\cite{Merabet:1993du}.

\subsection{Fermi-motion and the gluons in the EMC region}
 
Beside the scarce information on the gluon in the EMC region which 
can be obtained from global fits of nPDFs, 
some theoretical ideas have been put forward. For instance, 
it has been suggested~\cite{Merabet:1993du} that the stronger falloff at large-$x$ 
of the gluon PDF ($(1-x)^5$) vs the quark one ($(1-x)^3$) would have as consequence that the 
nPDF enhancement due to the Fermi-motion would manifest itself in gluon densities
down to smaller $x$  than  in quark ones. 

In practice, the Fermi-motion effect can simply be taken into account by convoluting
the nucleon PDF of a given flavour with a gaussian distribution 
encoding the effect of the Fermi momentum, $p_F$. 
Following~\cite{Merabet:1993du},
one has for the gluon density in a nucleus $A$:
\begin{equation}
g_{\rm A}(x)= \int_x^A dz\, N(z)\, \frac{1}{z} g\left(\frac{x}{z}\right), \hbox{with }N(z)=
\frac{1}{\sqrt{2 \pi \gamma}} \exp{\left(-\frac{(z-1)^2}{2\gamma}\right)},
\end{equation}
where $z$ corresponds to the light cone momentum fraction of the nucleon in the nucleus ($z= A p_N^+ / p_A^+$), 
$g(x)$ is the gluon distribution in the nucleon, 
$\gamma=\langle p_z^2 \rangle /M_N^2 = \frac{1}{5} p_F^2/M_N^2$ and $p_F=0.27$~GeV \cite{Bartke:2009zz}.

\begin{figure*}[htb!]
\begin{center}
{\includegraphics[height=0.4\textwidth]{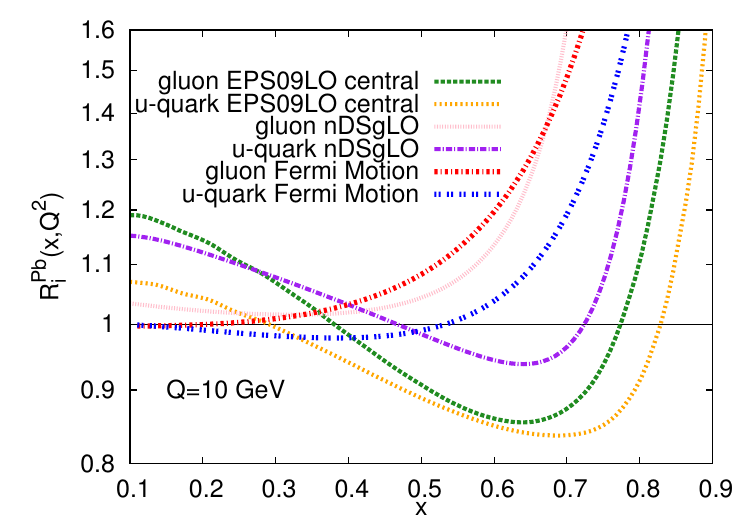}}
\end{center}\vspace*{-0.5cm}
\caption{Comparison between the nuclear modification of the gluon and $u$-quark distributions in 
a lead nucleus at large $x$ only due to the Fermi motion  following~\protect\cite{Merabet:1993du} 
(blue double dot, red dot-dash)
and those encoded in EPS09LO (green long-dash, yellow small dash) and nDSg (pink dot, purple dot-long dash).
}
\label{fig:FM-EMC}
\end{figure*}

\cf{fig:FM-EMC} shows our result for the $u$-quark and the gluon densities in a Pb nucleus 
modified by the Fermi motion\footnote{using, as initial proton PDF, 
the MSTW LO set~\cite{Martin:2009iq}.} compared to those from EPS09LO and from nDSg.
The impact of the Fermi motion is larger for the gluon than 
for the $u$-quark in the region of applicability of the model of~\cite{Merabet:1993du}. 
This arises uniquely from the different behaviour of the nucleon PDF at large $x$. 
One also observes that the gluon 
distribution is still enhanced at $x\simeq 0.4$ whereas the $u$-quark one is then nearly unmodified.
As can be seen on~\cf{fig:FM-EMC}, the trend of the nDSg fit
follows the behaviour expected from the Fermi motion. nDSg can thus be seen as a realisation of this simple model.
On the contrary, the 
EPS09LO fit shows a different behaviour since it incorporates additional
effects from antishadowing and from EMC suppression.

\subsection{Gluon saturation and $\Upsilon$ production at RHIC and LHC energies}
\label{subsecgluonsaturation}

In order to evaluate the saturation scale from which one expects effects
beyond collinear factorisation to be important   in $\Upsilon$ production in $pA$,
we have employed the following formula~\cite{Albacete:2009fh,Albacete:2010sy}:
\eqs{Q^2_{s \rm A}=A^{\frac{1}{3}}\times 0.2 \times \Big(\frac{x_0}{x}\Big)^\lambda \hbox{(in unit of GeV}^2),}
with $\lambda\sim 0.2\div 0.3$ and with $x_0=0.01$ which sets the minimum momentum fraction
below which one expects non-linear effects to be significant in the evolution of the parton distribution.
In the previous formula, we have set $x$ equal to $\langle x_2 \rangle$ computed with our 
Monte Carlo framework for a given rapidity. The values which we have obtained for $Q_{sA}$ are given 
in \ct{tab:Qs-RHIC} and \ct{tab:Qs-LHC} along
with the ratio of $Q_{sA}$ to the $\Upsilon$ mass. For the rapidities where 
$\langle x_2 \rangle$ was found to be above $x_0$, the saturation scale in a 
proton is of the order of $\Lambda_{QCD}$. Even in a large nucleus, it is still below 1~GeV.
In such a case, one thus does not expect any new phenomenon beyond collinear 
factorisation --\ie~beyond those encoded in the nPDFs-- to be significant in hard 
processes such as heavy-quark and quarkonium production.

\begin{table}[htb!]
\begin{center}\setlength{\arrayrulewidth}{1pt}
\begin{tabular}{cccc|cccc}
\hline\hline
 $y$ & $Q_{s\rm Au}$(GeV) & $\frac{Q_{s\rm Au}}{m_{\Upsilon}}$   & \quad& &$y$ & $Q_{s\rm Au}$(GeV) & $\frac{Q_{s\rm Au}}{m_{\Upsilon}}$ \\
\hline
-2.0 & $\lesssim 1$   & --  &   &    &  0.0 &   $\lesssim 1$     & -- \\
-1.5 & $\lesssim 1$   & --  &   &    & +1.5 &   $1.0 \div 1.1$     & $0.1$ \\
-1.0 & $\lesssim 1$   & -- &   &    & +2.0 &   $1.1 \div 1.2$     & $0.1$ \\
\hline\hline
\end{tabular}
\caption{Evaluation of the saturation scale  (for $x \leq x_0$) in the Au nucleus, $Q_{s\rm Au}$, 
and of its ratio to $m_\Upsilon$ for the kinematics of the $\Upsilon$ production at $\sqrt{s_{NN}}=200$~GeV
in $d$Au collisions as a function of the $\Upsilon$  rapidity (in the $d$Au CMS, \ie~the
laboratory system at RHIC).}\label{tab:Qs-RHIC}
\end{center}\vspace*{-0.5cm}
\end{table}

\begin{table}[htb!]
\begin{center}\setlength{\arrayrulewidth}{1pt}
\begin{tabular}{cccc|cccc}
\hline\hline
 $y$ & $Q_{s\rm Pb}$ (GeV) & $\frac{Q_{s\rm Au}}{m_{\Upsilon}}$  & \quad& & $y$  & $Q_{s\rm Pb}$(GeV) & $\frac{Q_{s\rm Pb}}{m_{\Upsilon}}$\\
\hline
-4.0 &  $\lesssim 1$  & -- &   &   & +2.0   &  $1.6\div 1.9$      & $0.2$ \\
-2.0 &  $\lesssim 1$  & -- &   &   & +4.0   &  $1.9 \div 2.5$      & $0.2-0.25$\\
0.0  & $1.3\div 1.4$    & 0.15 &   &                                   \\
\hline\hline
\end{tabular}
\caption{Same as \ct{tab:Qs-RHIC} at $\sqrt{s_{NN}}=5$ TeV
in $p$Pb collisions as a function of the $\Upsilon$ rapidity (in the CM).}\label{tab:Qs-LHC}
\end{center}\vspace*{-0.5cm}
\end{table}

Following the values  displayed on \ct{tab:Qs-RHIC} and \ct{tab:Qs-LHC}, one does not expect any
specific saturation effect on $\Upsilon$ production in $p(d)A$ collisions at RHIC and the LHC since the
saturation scale is always well below the typical energy scale of the process, namely $m_\Upsilon$ or even $m_b$.
In particular, the shadowing of the gluons as encoded in the nPDF fits based on the collinear factorisation 
should give a reliable account of the possible low-$x$ physics in the forward region.

\section{Results for \dAu\ collisions at RHIC}
Experimentally, the nuclear effects on $\Upsilon$ production in $d$Au is
 studied by measuring a {\it nuclear modification factor} $R_{d{\rm Au}}$, 
the ratio  of the yield in $d{\rm Au}$ collisions to the yield in $pp$ 
collisions at the same energy, times the average number of binary inelastic 
nucleon-nucleon collisions, $N_{coll}$, in a $d{\rm Au}$ collision: 
\eqs{R_{d{\rm Au}}=\frac{dN_{d{\rm Au}}}{\langle\Ncoll\rangle dN_{pp}}.}
Any nuclear 
effect affecting the \upsi\  production leads to  $R_{d{\rm Au}}\neq 1$.

From the data from STAR and PHENIX~\cite{Adare:2012bv,Reed:2010zzb},
only the rapidity dependence of $R_{d{\rm Au}}$ is known. For now, the 3 \upsi\ resonances are 
not resolved but are measured together. Since the nuclear absorption has 
to be small and since the nPDF effects are very likely similar for these 3 
states, we can safely consider them on the same footage. However, it is 
worth noting that a future measurement of $R_{d{\rm Au}}$ focusing only on 
$\Upsilon(1S)$ would be very precious to confirm this assumption.

\begin{figure*}[htb!]
\begin{center}
{\includegraphics[width=0.7\textwidth]{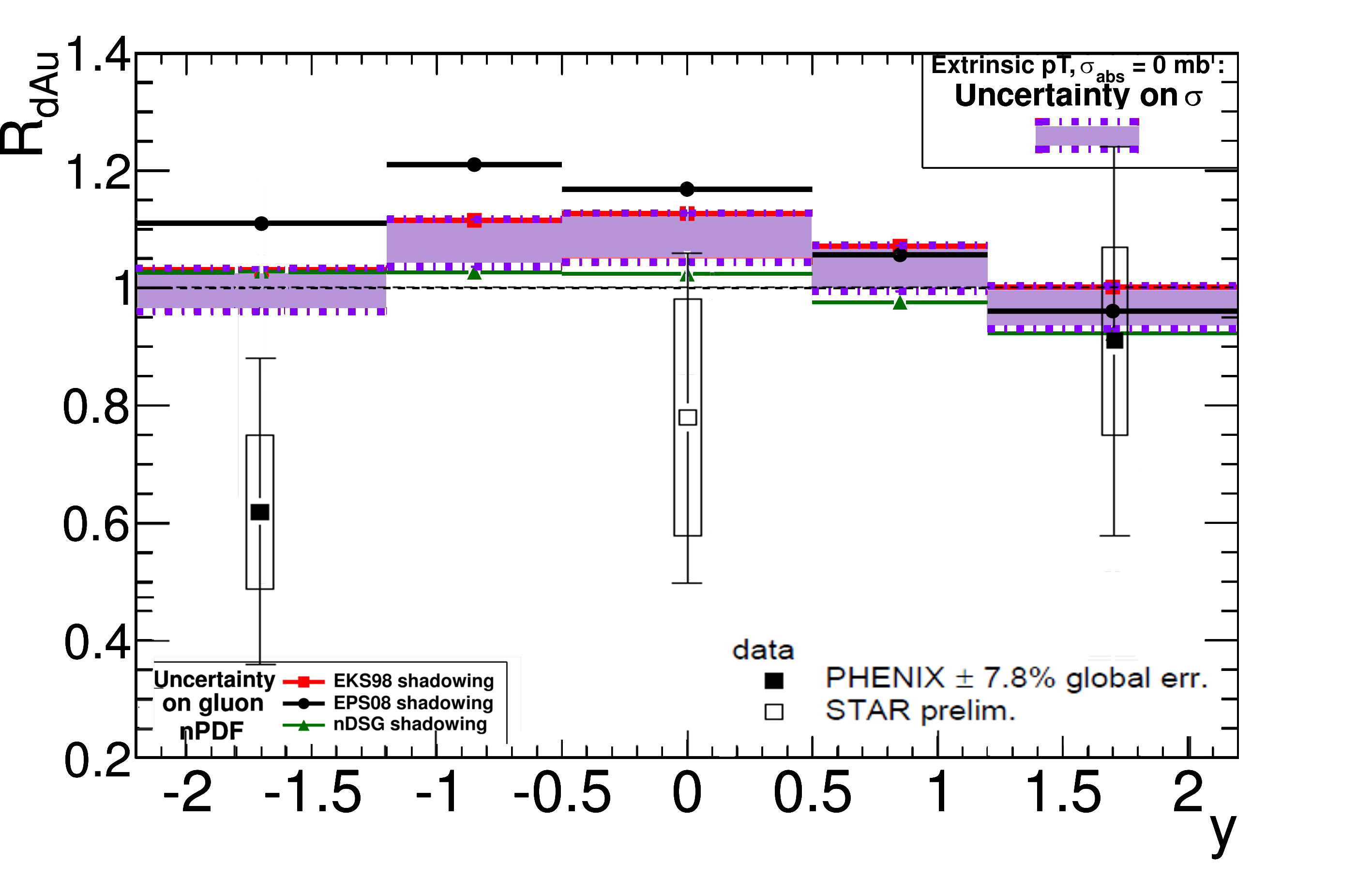}}
\\
\end{center}\vspace*{-0.5cm}
\caption{ 
Theoretical uncertainties on $R^{\Upsilon}_{d{\rm Au}}$ 
due to usual nPDFs (coloured lines) and $\sigma_\mathrm{eff}$ (purple band).
Data for $\Upsilon$ are from~\cite{Adare:2012bv,Reed:2010zzb}.
}
\label{fig:Rdau_vs_y_uncertainty}
\end{figure*}

\cf{fig:Rdau_vs_y_uncertainty} shows the uncertainties on $R_{dAu}$ vs $y$ due to
the lack of knowledge on the gluon nPDF and due to a variation of
 $\sigma_\mathrm{eff}$ between 0 and 1 mb.  The nuclear absorption (purple band) shows 
a very mild effect -- despite the likely exaggerated upper value we have taken. 
It is in any case insufficient to reproduce the available data in the backward rapidity region.
At mid $y$, the STAR data do not show any hint of antishadowing. 
In fact, the nDSg nPDF --without any antishadowing-- gives the best account of the data. 
If this was to be confirmed, this would be
an extremely important constrain for further fits of gluon nPDFs.
A measurement bearing on the sole $\Upsilon(1S)$ 
and possibly with a binning in $y$ which would allow one
to see if the suppression is getting even stronger for more negative $y$ or not, would therefore be invaluable.

At forward $y$, the shadowing encoded in the three usual nPDF fits we have used (EKS98, EPS08 and nDSg) 
agrees with the present PHENIX data.

\begin{figure*}[htb!]
\begin{center}
{\includegraphics[width=0.7\textwidth]{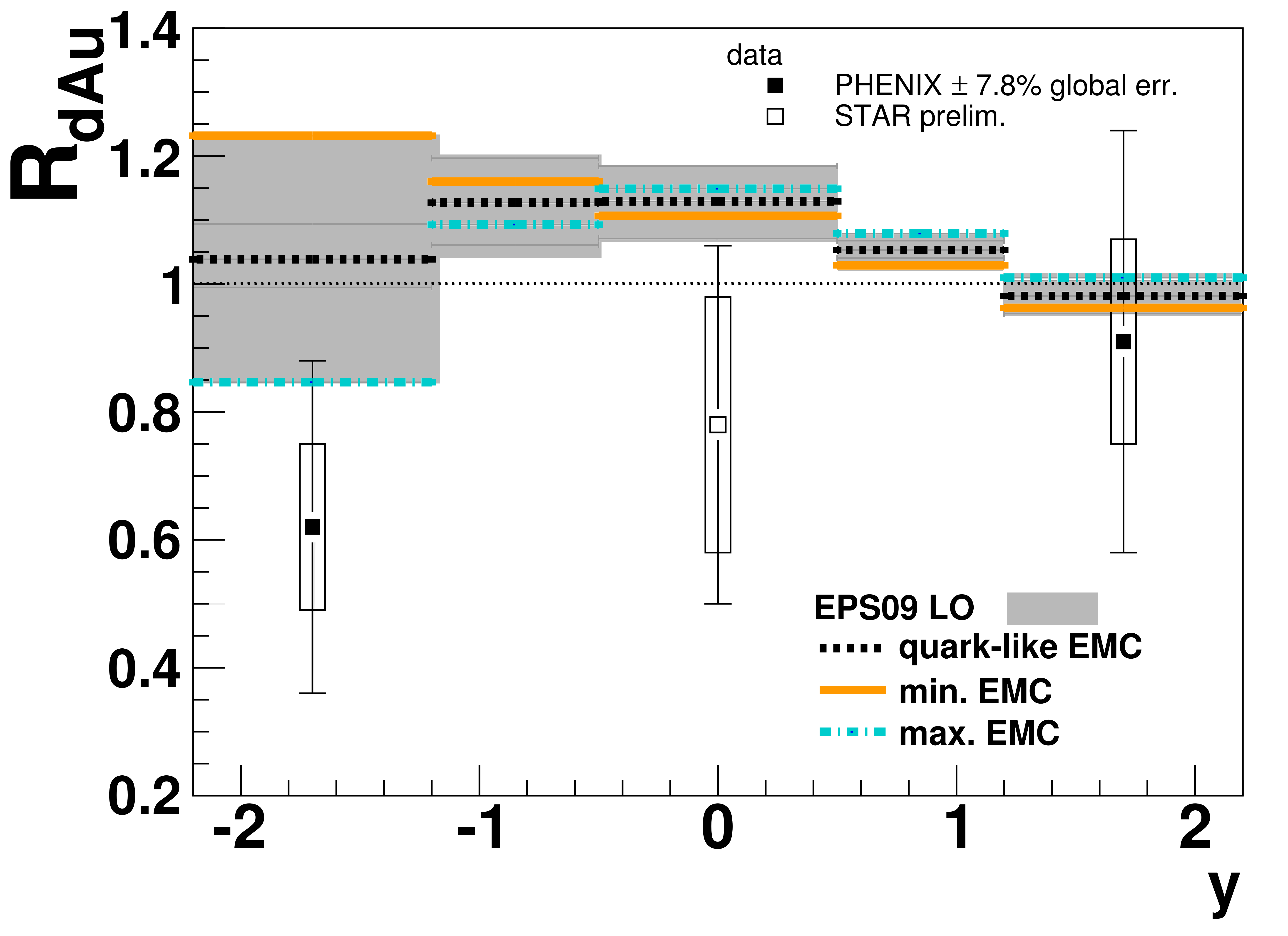}}
\end{center}\vspace*{-0.5cm}
\caption{ 
Effect of an absence and of an increase of the gluon EMC effect in EPS09LO on $R^{\Upsilon}_{d{\rm Au}}$.
Data for $\Upsilon$ are from ~\cite{Adare:2012bv,Reed:2010zzb}.
}
\label{fig:Rdau-EMC}
\end{figure*}

As one can see on \cf{fig:Rdau-EMC} for backward $y$, a gluon EMC effect stronger
 than that of quarks (dashed blue lines),  
such as the one provided by the EPS09 lower bound --or stronger--, 
would be provide a convincing account of the backward data.
A strong gluon EMC  effect is 
perfectly legitimate given the current knowledge of the gluon nPDF in this region.
The current experimental uncertainties are large 
and the observation of such a strong EMC gluonic effect would only be confirmed once we have
data with reduced errors. 
{The data however already visibly disfavour the case with no gluon EMC effect in EPS09LO (orange solid bars
on \cf{fig:Rdau-EMC})}
with an excess\footnote{Such excess of the gluon density in large nuclei would be in line with the expected enhancement due to the 
Fermi motion and the steep large-$x$ falloff of $g(x)$. } 
of the gluons all the way from the antishadowing region up to the Fermi motion one 
(see \cf{fig:EPS09-EMC-range}). 

In the absence of any antishadowing, let us also recall
that the behaviour of nDSg for $x > 0.1$ would be very close to what is expected from the Fermi motion effect
and would disagrees with the PHENIX value (\cf{fig:Rdau_vs_y_uncertainty}).
As already mentioned, the current data are not precise enough to draw further conclusions.

\section{Results for $p$Pb collisions at the LHC}
We have extended our study to the \upsi\ production for $p$Pb collisions at $\sqrt{s_{NN}}=5$~TeV. 
In \cf{figlhc}, the effect of the shadowing as encoded in EPS09 LO is shown to be important, in particular 
in the forward rapidity region.  Taking into account that in PbPb collisions at mid rapidities the 
shadowing effect is squared compared to $p$Pb, we can 
expect, in minimum bias PbPb collisions, a typical suppression due to shadowing of the order of 20 \%.

\begin{figure*}[htb!]
\begin{center}
\includegraphics[width=0.7\textwidth]{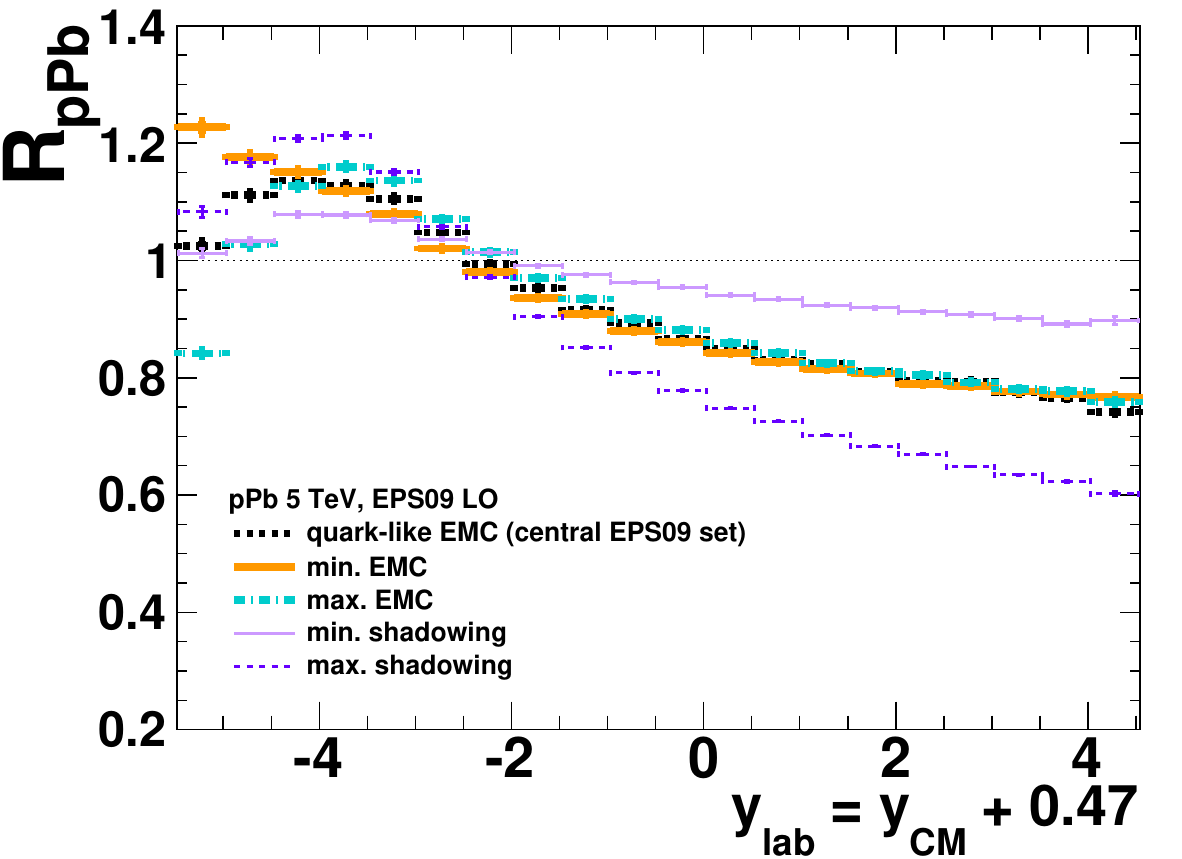}
\end{center}
\caption{
Effect of the nuclear modification of the PDF as encoded in EPS09 LO on $R^{\Upsilon}_{p{\rm Pb}}$ at 5 TeV taking
into account the rapidity shift between the CM and the laboratory system.}
\label{figlhc}
\end{figure*}

In addition, we suggest to analyse the forward-to-backward ratio of the nuclear modification factor, $R_{FB}(|y_{CM}|)\equiv R_{p{\rm Pb}}(y_{CM})/ R_{p{\rm Pb}}(-y_{CM})$ which has the advantage to be independent of a $pp$ reference and in which some systematic experimental uncertainties would cancel. Our predictions for such
a quantity are presented on \cf{figRFB} for the LHC kinematics. Such a forward-backward asymmetry is quite large (up to 45\%) and should be mesurable.

\begin{figure*}[htb!]
\begin{center}
\includegraphics[width=0.7\textwidth]{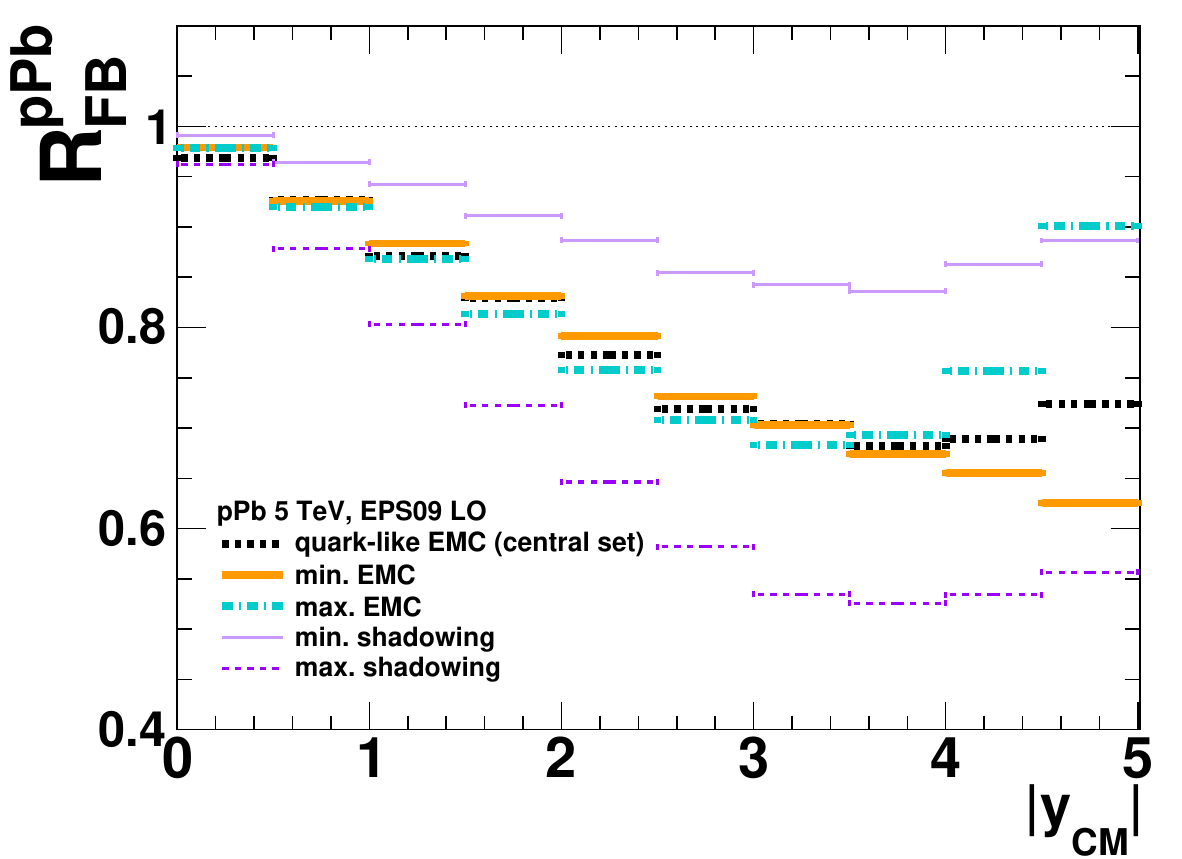}
\end{center}
\caption{
Forward-to-backward  nuclear modification factor, $R_{FB}(|y_{CM}|)\equiv R_{p{\rm Pb}}(y_{CM})/ R_{p{\rm Pb}}(-y_{CM})$, for
$\Upsilon$ production in $p$Pb collisions at 5 TeV as a function of the absolute value of $y_{CM}$}
\label{figRFB}
\end{figure*}

Provided that the only nuclear effect on the $\Upsilon$ production at the LHC would come from the modification of the PDFs, 
it is interesting to figure out how similar the suppression/enhancement of the 3 
$\Upsilon$ states would be. Two effects can enter: a shift in $x_2$ and a difference in 
the scale of the process, $Q^2$. We have checked, using EPS09LO~\cite{Eskola:2009uj}, EPS09NLO~\cite{Eskola:2009uj}, 
nDSgLO~\cite{deFlorian:2003qf}, HKN04LO~\cite{Hirai:2004wq} and with conservative choices of the $Q^2$, 
that the nuclear modification factors of $\Upsilon(nS)$
are expected to be equal to a precision of 2\%.

Yet, new effects might come into play. One of these is the fractional energy loss discussed in the next section,
which shows a magnitude inversely proportional to the mass state. 
This is, however, not sufficient to produce any visible difference within the $R_{p{\rm Pb}}$ of
the 3 $\Upsilon$ states. In fact, the difference would also be of the order of a few percents.
 As a reminder, conventional nuclear absorption can neither be an effect impacting differently on
the 3 states, simply  because only a single type of pre-resonant state propagates in the nuclear matter at the LHC.

Unfortunately, as long as one does not have at our disposal a $pp$ reference at 5 TeV, 
the uncertainties in the normalisation of the nuclear modification factor  would of course not allow
to reach precision of a few percents. It will thus be expedient to analyse differences in central-to-peripheral 
nuclear modification factors, $R_{CP}$ or forward-to-backward  nuclear modification
factors, $R_{FB}$ for 3 $\Upsilon$ states.

\section{Energy loss and production mechanism}

It has recently~\cite{Arleo:2010rb} been pointed out that, in the case 
of small angle quarkonium production, the spectrum of the gluon radiation induced by the nuclear 
medium can scale as the quarkonium energy $E$. The main reason for such a fact is that the gluon radiation  
 arises from large formation times $t_f \gg L$ and it is not subject to the 
bound derived in~\cite{Brodsky:1992nq} preventing any parton energy loss to scale 
like the projectile energy and thus to impact quarkonium production at RHIC and LHC energies.

This fractional energy loss may however only act on octet-like mechanisms\footnote{Along the same lines, it
would not have any effect on DY pair production for instance.} where the heavy-quark 
pair is produced at short distances in a colour-octet state and where the latter has thus a long enough time to radiate. 
This is a priori so in the Colour-Evaporation Model (CEM) and the Color-Octet Mechanism (COM) 
(see~\cite{Lansberg:2006dh,Brambilla:2010cs}). 
For singlet-like mechanisms --such as the CSM which is favoured by the low $P_T$ $pp$ data \cite{Lansberg:2012ta}--
 for which the heavy-quark pair is produced at short distances (or at short production times, $t_{\rm prod}$) in 
a colour-singlet state, such an energy loss is not expected to occur --at least as a fractional energy loss, scaling in $E$.

Following \cite{Arleo:2010rb}, one finds  that for forward angle $\Upsilon$ production in $pA$ collisions
via a long-lived color-octet pair, the fraction of medium-induced radiated energy is given by 
$\Delta E/E\simeq N_c \alpha_s \sqrt{\Delta\langle p_T^2\rangle}/M_T$, 
where  $\Delta\langle p_T^2\rangle$ is the broadening of the {radiated} gluon from the proton and
$M_T$ is the transverse mass of the final-state coloured object. 
$\Delta\langle p_T^2\rangle$ can be indirectly fit from the data as in~\cite{Arleo:2012hn}. Its
size can also be guessed from the \upsi\ broadening,
$\Delta\langle {P^\Upsilon_T}^2\rangle\equiv \langle {P^\Upsilon_T}^2\rangle (A)-\langle {P^\Upsilon_T}^2\rangle (^2$H), proportional 
to the length of nuclear matter seen by {the} incoming parton. This broadening of ${P^\Upsilon_T}^2$
is consistent with a dependence on $A^{1/3}$, {as} expected in multiple scattering models. 
Taking the E772 value\footnote{We note a discrepancy between the published
value of Ref.~\cite{Alde:1991sw} and that subsequently published in a review by members of 
E772~\cite{McGaughey:1999mq}. We prefer to use the latter since ``[T]his difference was due to the use of earlier parameterisations 
of the $P_T$ dependence, derived by E605 for $p$Cu collisions, rather than the new function based on 
$p\, ^2$H data''~\cite{McGaughey-unp}.} with W target~\cite{McGaughey:1999mq,Alde:1991sw}, 
$\Delta\langle {P^\Upsilon_T}^2\rangle=0.410$~GeV$^2$, one can thus write $\Delta\langle {P^\Upsilon_T}^2\rangle= 0.072$~GeV$^2$ $A^{1/3}$.

For $N_c=3$ and $\alpha_s=0.2$, we estimate a maximum loss of the order of  $\Delta E^{max}/E\sim 4\%$. 
It is important to note that $\chi_{b2}$ production at LO does not involve the emission of 
a coloured heavy quark pair, both in the CSM and the COM. Any feed-down from it would not be affected
by such an energy loss.
For forward $\Upsilon$ produced at RHIC, this energy loss implies 
a suppression of the order 10-15 \% when implemented in the PDFs
(see \cf{fig:Rdau-Eloss}).

\begin{figure*}[htb!]
\begin{center}
{\includegraphics[width=0.7\textwidth]{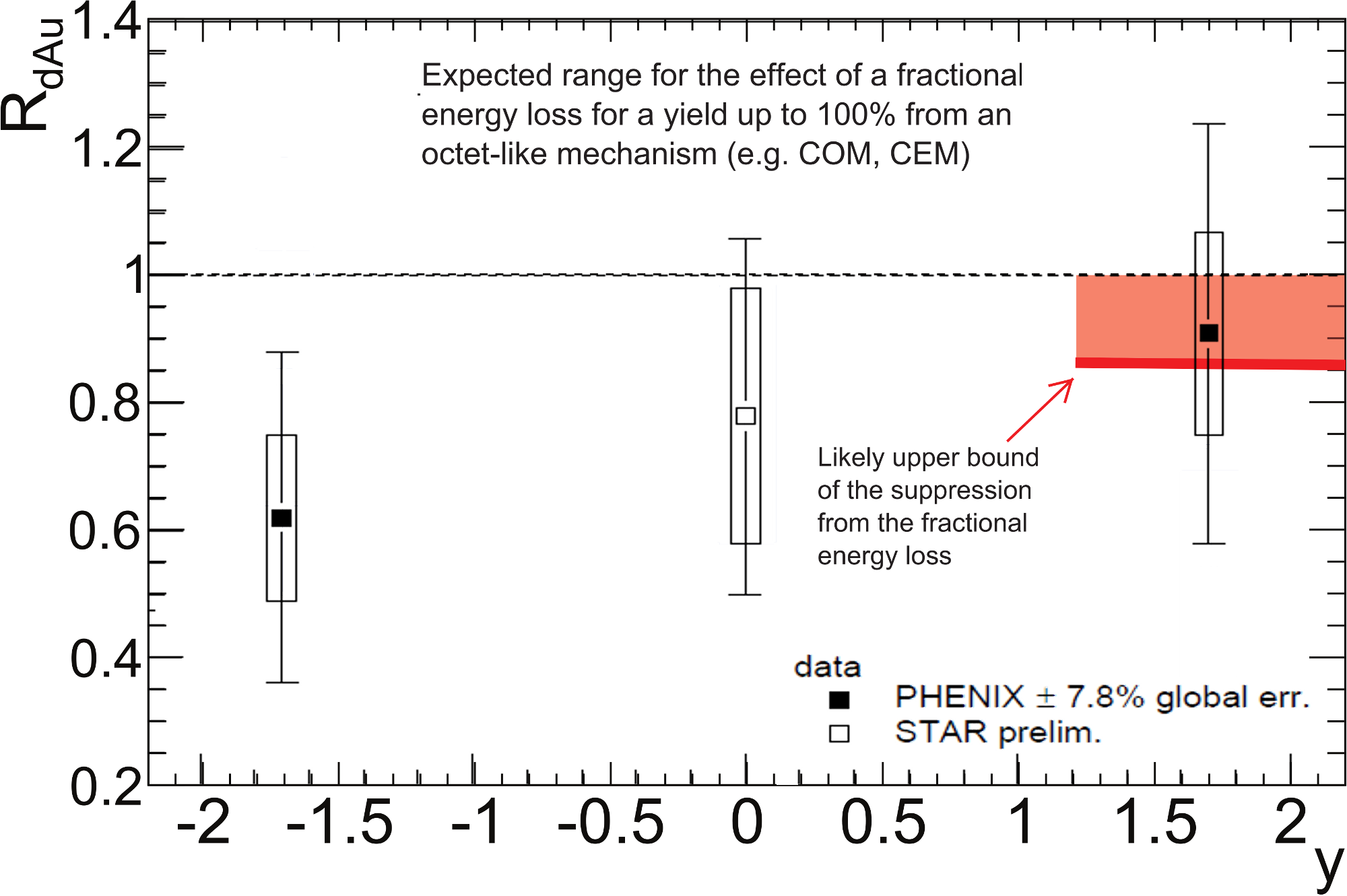}}
\end{center}\vspace*{-0.5cm}
\caption{ 
Expected range for the energy loss on $R^{\Upsilon}_{d{\rm Au}}$  for a production mechanism where
the heavy-quark remains in a coloured state for a long time.
Data for $\Upsilon$ are from\cite{Adare:2012bv,Reed:2010zzb}.
.
}
\label{fig:Rdau-Eloss}
\end{figure*}

Moreover, this suppression by an energy loss undergone by the heavy-quark pair before its
hadronisation is in apparent conflict with the convincing account of the $P_T$-integrated yields by the CSM
approach --see for instance~\cite{Lansberg:2012ta} for a recent account of the comparison between the CSM prediction and 
the results at RHIC, the Tevatron and the LHC. It is worth mentioning that the computation in the CSM does not 
involve any adjustable parameter
and that any contribution from colour-octet transitions would be additive. We would face 
an excess of the theoretical prediction in $pp$ as compared to the existing measurements if we were to
invoke a significant fraction of the yield to be suppressed in $pA$ since from a colour-octet like production
mechanism. 

In fact, for the time being, there is no up-to-date study of $\Upsilon$ production at low $P_T$ (below 5~GeV) 
incorporating in a consistent way the contributions from the leading colour-octet transitions. 
The most recent $\Upsilon$ analysis with CO at NLO did not even attempt to consider $P_T$'s 
below 7~GeV~\cite{Wang:2012is}.  In any case, 
to clear up the situation, at least at RHIC energy, it is of paramount importance to have at our disposal
experimental measurements of the $\Upsilon(1S)$ yield integrated and differential in $P_T$ as well as in rapidity.
Hopefully, such measurements would be available soon with the data accumulated at RHIC.

\section{Conclusion}
We have investigated the \upsi\ suppression in \dAu\ collisions at $\sqrt{s_{NN}}=200$~GeV at RHIC
and in $p$Pb collisions at $\sqrt{s_{NN}}=5$ TeV at the LHC. The  rapidities covered by the RHIC 
experiments allow for a unique study of cold nuclear matter effects and revealed unexpected features presented here. 
Backward rapidities correspond to  the largest $x_2$, above 0.2, where one can expect an EMC suppression -- at least for quark PDFs.
For mid and slightly backward rapidities, one expects anti-shadowing, \ie\ an excess of partons inducing an excess of $\Upsilon$. 
Finally, the forward domain -- where $x_2$ becomes small --  should be subject to parton shadowing, giving  
a reduction of the yield.

We have first argued that, for bottomonia, as opposed to charmonia, the survival probability for the pre-resonant 
state to escape  a large nucleus should be large and close to unity. In turn, the use of a 
small effective absorption cross section is mandatory. 

Then, we have discussed the different effects which can be expected from the modification of the gluon densities in
a nucleon bound in a heavy nucleus, namely from high $x$ to low $x$: the Fermi motion, the EMC effect, the antishadowing, 
the shadowing and then the saturation.

We have confronted these expectations with the existing data and 
our findings were as follows: in the most backward region, studied by PHENIX, 
the suppression of the $\upsi$ yield in the \dAu\ data may be a first hint
of an EMC suppression of the gluon PDF, possibly stronger than the quark one. It is one of the
first observations of such an effect, whose quantitative understanding may
in the future provide us with fundamental information on the internal dynamics of
heavy nuclei such as those studied at RHIC, especially if the EMC gluon effect is
stronger than the quark one. In the central rapidity region, the existing
STAR data  does not exhibit any excess which would pin down anti-shadowing.

In addition, due to the large scale set in by the $\upsi$ mass, 
shadowing is found to be small and it reproduces the present PHENIX data~\cite{Adare:2012bv}
 at $y>0$. As we discussed, additional effects sometimes claimed to apply to $J/\psi$ production 
--such as saturation or fractional energy loss-- are not relevant here and, in fact, are not needed.
We have indeed demonstrated that --at variance with the $J/\psi$ case~\cite{Ferreiro:2009ur,Ferreiro:2012sy}-- 
gluon shadowing has, for any nPDF fit, a small effect at the $x_2$ and $Q^2$ of forward-$\Upsilon$ production 
at RHIC.

We have finally concluded that additional data at RHIC energies both in $pp$ and $d$Au collisions 
are eagerly awaited as well as the forthcoming LHC data in $p$Pb collisions at 5 TeV.

\section*{Acknowledgements}
We are grateful to J. Albacete, F.Arleo, N. Armesto, R. Arnaldi, S.J. Brodsky, 
B.Z. Kopeliovich, M. Leitch, S. Peign\'e and C. Salgado, for useful
discussions.
This work is supported in part by 
Ministerios de Educacion y 
Ciencia of Spain and IN2P3 of France (AIC-D-2011-0740) and by the ReteQuarkonii Networking of the EU I3 Hadron
Physics 2 program. N.M. thanks the F.R.S.-FNRS (Belgium) for financial support.

\end{document}